\documentstyle[preprint,aps,prd]{revtex}
\draft
\begin{document}
\title
{Gaussian Time-Dependent Variational Principle for  Bosons
Contact Interaction in one dimension}
\author
{Arthur K. Kerman}
\address{Center for Theoretical Physics, Laboratory for Nuclear Science}
\address{and Department of Physics}
\address{Massachusetts Institute of Technology}
\address{Cambridge, MA 02139}
\author
{Paolo Tommasini}
\address{Institute for Theoretical Atomic and Molecular Physics}
\address{Harvard-Smithsonian Center for Astrophysics}
\address{Cambridge, MA 02138}
\date{\today}
\maketitle
\begin{abstract}
We investigate the Dirac time-dependent variational method using 
a Gaussian trial functional for an infinite one dimensional 
system of Bosons interacting through a repulsive contact 
interaction. The method produces a set of non-linear time 
dependent equations for the variational parameters. By 
solving the static equations we have calculated the ground 
state energy per particle. We have also considered small 
oscillations about equilibrium and obtain  mode  equations 
which lead us to a gapless dispersion relation. 
The existence of an exact numerical solution for the ground state energy
and excitations obtained by Lieb allow us to compare with the Gaussian results. We
can also, as  the system becomes less dilute, see the improvement of the results
as compared with the Bogoliubov scheme.

\end{abstract}
\pacs{PACS numbers: 03.75.Fi, 05.30.Jp, 32.80Pj, 67.90.+z}
\narrowtext
\section{Introduction}
Recently  Bose-Einstein condensation  in atomic traps  was achieved \cite{AE}-\cite{KM1} in a dilute
regime ($a^{3} n << 1$). This is in  contrast with  the Helium 4 regime where
a much higher density leads to ($a^{3} n \approx 1$). For the high density
regime there is a large  "depletion"  ($\approx 90\%$) \cite{OO} where as for
the dilute regime it is  very small ($\approx 1 \% $) \cite{TR}.

Theoretically these new experiments have been  described very successfully
with  mean  field theory without
quantum fluctuations using the Gross Pitaevskii equation \cite{PT}  or 
 the
Bogoliubov scheme \cite{TR}.
The Gaussian self-consistent approximation  presented here should  
be useful
at in  intermediate regime. Unfortunately due to three body recombinations
there seems to be a limit \cite{HG} for
increasing the number of particles in the system. Recently it has been
pointed out \cite{VR} that
 using a
strong magnetic field  it would be possible to make the system effectively 
less dilute by drastically changing the scattering length.
This  makes
the comparison between self-consistent results and  dilute theories very
important.
In this context the one dimensional delta function   case can produce some insight because a
contact interaction can be used in the self-consistent theory in contrast with
the 3 dimensional case \cite{HP} - \cite{PP}. The existence of
  an exact solution for the ground state energy and for particle
and hole excitations makes the comparison very intersting,  provide that we
understand how the separated particle and hole excitations of the exact solution \cite{LB2} are
connected to the  particle-hole  excitations given by the approximate methods.

The objective of this paper is to exhibit the most general way of obtaining
time dependent equations of motion in the Gaussian approximation \cite{KK}. This
will lead to the so called generalized RPA, when one
examines infinitesimal oscillations about equilibrium.
The static solution in the uniform case
 can be obtained using several other methods
\cite{GG} - \cite{HP} leading to a gap in the quasi-boson energy.
We show here that  the time dependent RPA equations lead to a gapless mode.
In fact this {\bf must}  happen because  particle number
 conservation symmetry is broken in the static solution, so the zero gap
 is exactly the
  associated Goldstone mode. This discussion can be seen as an alternative to
the
functional derivative \cite{HM}, \cite{SS} method in the Girardeau-Arnowitt
\cite{GR}
approximation.

The Bogoliubov
scheme, for a dilute or weak interacting system,  can  be obtained by
a particular truncation of the
Gaussian results. So that  we can  compare the Gaussian variational results ,
the dilute Bogoliubov  scheme and the exact solution, for the particular 
case under discussion here.

The structure of this paper is as follows. Section 2 reviews the time-dependent
 variational principle and the canonical nature of the equations of motion
arising from it. In section 3 we specialize to the one dimensional uniform 
case end examine the ground state energy and the excitations for both the
approximate
methods and the exact solution. Section 4 contains our numerical solutions and
conclusions

\section{General Formalism}
In this section we shall review some of the   results of the
 Time-dependent
Variational Principle {\cite{KK}},{\cite{KL}} and show how it can be 
implemented in the
non-relativistic case. First we define an effective action 
functional for the
time-dependent quantum system
\begin{equation}
\label{1}
S = \int L(t) dt = \int dt \langle \Psi,t | (i \partial_{t}
- \hat{H}) | \Psi, t \rangle,
\end{equation}
\noindent where $| \Psi,t \rangle$ is the quantum state of 
the system and
$\hat{H}$ is the Hamiltonian of the theory. For a system of
non-relativistic
interacting Bosons we have [we use the notation : $\int_{\bf
 x} = \int d^{3}x$]
\begin{equation}
\label{2}
\hat{H} = \int_{{\bf x},{\bf y}}  \hat{\psi}({\bf x})^{\dag}
 h({\bf x},{\bf y}) \hat{\psi}({\bf y})  + \frac{1}{2}\int_{
{\bf x},{\bf y}}
\hat{\psi}({\bf y})^{\dag}\hat{\psi}({\bf x})^{\dag} V({\bf
x} - {\bf y})
\hat{\psi}({\bf x}) \hat{\psi}({\bf y}),
\end{equation}
\noindent where the one body Hamiltonian $h({\bf x},{\bf y})
$ may include a one
 body external potential. The creation and destruction operators $\hat{\psi}^{\dag}$ and 
$\hat{\psi}$
can be written in the form
\begin{eqnarray}
\label{3}
\hat{\psi}({\bf x}) &=& \frac{1}{\sqrt{2}} \left[ \hat{\phi}
({\bf x}) + i \hat{\pi}({\bf x}) \right] \nonumber \\
\\
\label{4}
\hat{\psi}({\bf x})^{\dag} &=& \frac{1}{\sqrt{2}} \left[ 
\hat{\phi}({\bf x}) -i \hat{\pi}({\bf x}) \right] \nonumber
\end{eqnarray}
\noindent where $\hat{\phi}({\bf x})$ is the field operator
and
$\hat{\pi}({\bf x})$ the canonical field momemtum

We can obtain the time dependent Schr\"odinger equation by 
requiring that $S$
is stationary, supplemented by appropriate boundary 
conditions,
 under the most general variation of $| \Psi ,t \rangle$.
The variational scheme is implemented by chosing  a trial wave 
functional
describing the
system. Working in the functional Shr\"odinger picture we 
replace the abstract
state $| \Psi ,t \rangle$ by a wave functional of the field
$\phi'({\bf x})$
\begin{equation}
\label{5}
| \Psi,t \rangle \rightarrow \Psi [ \phi',t].
\end{equation}
\noindent The action of the operators $\hat{\phi}({\bf x})$
and the canonical momentum
$\hat{\pi}({\bf x})$ are realized respectively by
\begin{eqnarray}
\label{6}
\hat{\phi}({\bf x}) |\Psi,t \rangle & \rightarrow & \phi'({
\bf x}) \Psi[\phi',t]  \nonumber \\
\\
\label{7}
\hat{\pi}({\bf x}) | \Psi,t \rangle & \rightarrow & - i 
\frac{\delta}{\delta
\phi'({\bf x})} \Psi[\phi',t]. \nonumber
\end{eqnarray}
\noindent The mean value of any operator is calculated by 
the functional integral
\begin{equation}
\label{8}
\langle \Psi,t |{ \cal O} | \Psi , t \rangle = \int ({\cal D
} \phi') \Psi^{\ast}
[\phi' , t] {\cal O} \Psi[\phi' , t],
\end{equation}
\noindent where $\Psi$ is normalized to unity. The Gaussian
approximation
consists in taking a Gaussian trial wave functional in its 
most general
parametrization
\begin{eqnarray}
\label{9}
\Psi[\phi',t] &=& N \exp\left\{ - \int_{{\bf x},{\bf y}}  
\delta \phi'({\bf x},t) \left[\frac{G^{-1}({\bf x},{\bf y},t)}
{4} - i \Sigma({\bf x},{\bf y},t) \right] \delta \phi'({\bf
y},t) \right. \nonumber \\
&& + \left.i \int_{\bf x}  \pi({\bf x},t)\delta\phi'({\bf x}
,t)
 \right\},
\end{eqnarray}
\noindent
with $\delta \phi'({\bf x},t) = \phi'({\bf x}) - \phi({\bf x
},t) $. Due to the fact that the Hamiltonian commutes with 
the number of particles
$\hat{N} = \int_{\bf x} 
\hat{\psi}^{\dag} ({\bf x})  \hat{\psi}({\bf x})$ i.e.
\begin{equation}
\label{10}
[\hat{H},\hat{N}] =0
\end{equation}
\noindent we can actually define a more general trial functional
\begin{equation}
\label{11}
| \Psi',t \rangle  = e^{-i \hat{N} \theta(t)}  |\Psi,t 
\rangle,
\end{equation}
\noindent where $\theta(t)$ is another variational parameter
 introduced because of this continuous symmetry.
Thus  our variational parameters are $\phi({\bf x},t)$,
$\pi({\bf x},t)$, $\theta(t)$,$G({\bf x},{\bf y},t)$ and
$\Sigma({\bf x},{\bf y},t)$ ,with  $G$ and $\Sigma$ being  
real symmetric matrices.
These quantities are related to the following mean-values:
\begin{eqnarray}
\label{12}
\langle \Psi' , t | \hat{\phi}({\bf x}) | \Psi' , t 
\rangle
 & = & \phi({\bf x},t) \nonumber \\
\label{13}
\langle \Psi' , t | \hat{\pi}({\bf x}) | \Psi' , t \rangle
 & = & \pi({
\bf x},t) \\
\label{14}
\langle \Psi' , t | \hat{\phi}({\bf x})  \hat{\phi}({\bf y
}) | \Psi' , t \rangle & = & G({\bf x},{\bf y},t) + \phi({
\bf x},t) \phi({\bf y},t) \nonumber \\
\label{15}
\langle \Psi' , t |i \frac{\delta}{\delta t}  | \Psi' , t 
\rangle  & = &
\int_{{\bf x},{\bf y}}  \Sigma({\bf x},{\bf y},t) \dot{G}({
\bf y},{\bf x},t) + \int_{\bf x}  \pi({\bf x},t) \dot{\phi}({
\bf x},t)  \nonumber \\
&&+{\cal N} \dot{\theta}(t) + \mbox{ total time derivatives}.
\end{eqnarray}
\noindent We may
 ignore the total time derivatives because they do not
contribute to the equations of motion. If now we write the 
action we will get
\begin{equation}
\label{16}
S = \int dt \left(\int_{\bf x} \pi({\bf x},t)\dot{\phi}({\bf
 x},t) +
\int_{{\bf x},{\bf y}} \Sigma({\bf x},{\bf y},t) \dot{G}({
\bf y},{\bf x},t) +
{\cal N} \dot{\theta} (t) - {\cal H} \right),
\end{equation}
\noindent where
\begin{equation}
\label{17}
{
\cal H} =\langle \Psi' , t | \hat{H}| \Psi' , t \rangle
\end{equation}
\noindent and
\begin{equation}
{\cal N} = \langle \Psi' , t | \hat{N}| \Psi' , t \rangle.
\end{equation}
\noindent From (\ref{16}) we see that $({\cal N},\theta)$, $
(\pi,\phi)$ and
$(\Sigma,G)$ are canonical pairs.
Because of the symmetry ${\cal H}$ has no dependence on $
\theta$ and it
follows that
$\dot{N} =0$ and $\dot{\theta}(t) =$ constant $ \equiv \mu $
. We can now write
the remaining Hamilton equations,
\begin{eqnarray}
\label{18}
\dot{\phi}({\bf x},t) &=& \frac{\delta( {\cal H} - \mu {\cal
 N})}{\delta \pi ({\bf x},t)}, \nonumber \\
\label{19}
\dot{\pi}({\bf x},t) &=& - \frac{\delta( {\cal H} - \mu {
\cal N})}{\delta \phi ({\bf x},t)}, \\
\label{20}
\dot{G}({\bf x},{\bf y},t) &=& \frac{\delta( {\cal H} - \mu
{\cal N})}{\delta
\Sigma({\bf x},{\bf y},t)}, \nonumber \\
\label{21}
\dot{\Sigma}({\bf x},{\bf y},t) &=&- \frac{\delta( {\cal H}
- \mu {\cal N})}{\delta
G({\bf x},{\bf y},t)}. \nonumber 
\end{eqnarray}
\noindent For convenience we introduce
\begin{equation}
\label{22}
\psi({\bf x},t) \equiv \langle \hat{\psi}({\bf x}) \rangle =
 \frac{\phi({\bf x},t) + i \pi({\bf x},t)}{\sqrt{2}},
\end{equation}
\noindent so that the equations for $\phi$ and $\pi$   become
\begin{equation}
\label{23}
\imath \dot{\psi}({\bf x},t) = \frac{\delta( {\cal H} - 
\mu {\cal N})}{\delta \psi^{\ast}({\bf x},t)}
\end{equation}
To obtain ${\cal H} -\mu {\cal N}$ we have to compute
\begin{equation}
{\cal H} -\mu {\cal N} = \int ({\cal D} \phi') \Psi^{\ast}[\phi',t] \left[ \hat{H} - \mu \hat{N} \right
] \Psi[\phi',t]
\end{equation}
\noindent using (\ref{3}) and (\ref{6}) we have

\begin{eqnarray}
\label{23aa}
&&{\cal H} -\mu {\cal N} = \int ({\cal D} \phi') \Psi^{\ast}[\phi',t] \left(\phi'({\bf x}) -
\frac{\delta }{\delta \phi'({\bf x})}\right) h({\bf x},{\bf y}) \left(\phi'({\bf y}) +
\frac{\delta }{\delta \phi'({\bf y})}\right) \Psi[\phi'] \nonumber \\
&&+ \int_{{\bf x},{\bf y}} ({\cal D} \phi')\Psi^{\ast}[\phi'] \left(\phi'({\bf x}) - \frac{\delta }{\
\delta \phi'({\bf x})}\right)
\left(\phi'({\bf y}) - \frac{\delta }{\delta \phi'({\bf y})}\right)V({\bf x}-{\bf y}) \nonumber \\
&& \times \left(\phi'({\bf x})+ \frac{\delta }{\delta \phi'({\bf x})}\right) \left(\phi'({\bf y} +
\frac{\delta }{\delta \phi'({\bf y})}\right) \Psi[\phi']
\end{eqnarray}

All the functional integrals can be easily computed using an
aditional source term (Appendix A) leading to

\begin{eqnarray}
\label{26}
{\cal H} - \mu {\cal N} = &\int_{{\bf x},{\bf y}}& \left\{\left[h({
\bf x},{\bf y}) - \mu \delta({\bf x}-{\bf y})  \right]  \rho({\bf x},{\bf y},t)
+ \frac{1}{2} V({\bf x}-{\bf y})
|\psi({\bf x},t)|^{2}  |\psi({\bf y},t)|^{2}
 \right\}\nonumber \\
+ \frac{1}{2} & \int_{{\bf x},{\bf y}}& V({\bf x}-{\bf y}) \left[ 
R({\bf y},{\bf x},t) R({\bf x},{\bf y},t) + 
R({\bf x},{\bf x},t) R({\bf y },{\bf y},t) +
D^{\ast} ({\bf x},{\bf y},t) D({\bf x},{\bf y},t) \right] \nonumber \\
+  &\int_{{\bf x},{\bf y}}& V({\bf x}-{\bf y}) \left[
  \frac{1}{2} \psi^{\ast}({\bf x},t)\psi({\bf y},t) R({\bf x},{\bf y},t) 
+  \frac{1}{2} \psi^{\ast}({\bf y},t)\psi({\bf x},t) R({\bf y },{\bf x},t)
+  |\psi({\bf x},t)|^{2} R({\bf y },{\bf y},t) \right] \nonumber \\
- \frac{1}{2} &\int_{{\bf x},{\bf y}}&  V({\bf x}-{\bf y}) \left[
\psi({\bf x},t) \psi({\bf y},t) D^{\ast}({\bf x},{\bf y},t) +
\psi^{\ast}({\bf x},t) \psi^{\ast}({\bf y},t) D({\bf x},{\bf y},t)
\right] 
\end{eqnarray}
\noindent and 
\begin{eqnarray}
\label{27}
\rho({\bf x},{\bf y},t) &=& \langle \psi^{\dag}({\bf x}) 
\psi({\bf y}) \rangle= \psi^{\ast}({\bf x},t) \psi({\bf y},t)
+ R({\bf x},{\bf y},t) \nonumber \\
\label{28}
\\
\Delta({\bf x},{\bf y},t) &=&-\langle \psi({\bf x}) \psi({\bf
 y}) \rangle = - \psi({\bf x},t) \psi({\bf y},t) + D({\bf x}
,{\bf y},t) \nonumber 
\end{eqnarray}
\noindent with
\begin{eqnarray}
\label{29}
R({\bf x},{\bf y},t) &=& \frac{1}{2} \left[ \frac{G^{-1}({
\bf x},{\bf y},t)}
{4} + G({\bf x},{\bf y},t) - \delta({\bf x}-{\bf y})
\right]  
+2 \int_{{\bf w},{\bf z}}  \Sigma({\bf x},{\bf w},t) G({
\bf w},{\bf z},t)
\Sigma({\bf z},{\bf y},t)  \nonumber \\
&& + i \int_{\bf z} \left[G({\bf x},{\bf z},t) \Sigma({\bf z},{\bf y},t) - \Sigma({\bf x},{\bf z},t) 
G({\bf z},{\bf y},t) \right] \nonumber \\
\\
\label{30}
D({\bf x},{\bf y},t) &=& \frac{1}{2} \left[ \frac{G^{-1}({
\bf x},{\bf y},t)}{4} -G({\bf x},{\bf y},t) \right]
+  2 \int_{{\bf w},{\bf z}}
 \Sigma({\bf x},{\bf w},t) G({\bf w},{\bf z},t) \Sigma({\bf
z},{\bf y},t) \nonumber   \\
&& -i \int_{{\bf z}} \left[\Sigma({\bf x},{\bf z},t) G({\bf z},{\bf y},t) +
 G({\bf x},{\bf z},t)\Sigma({\bf z},{\bf y},t)\right], \nonumber
\end{eqnarray}
\noindent because of (\ref{12}). 
It is easy to check that in terms of $R$ and $D$ the mean value ${\cal H} - \mu {\cal N}$ corresponds
to the standard mean-field factorization \cite{HP},\cite{GG}.
We note that the
 density gets contributions from the condensate field
$\psi$ as well as from the fluctuations ($G$,$\Sigma$). The
contribution from
$\psi^{\ast} \psi$ is the condensate density.
So that the term with 4 $\psi$'s can be interpreted as the condensate
self-interaction. The
interaction of particles not in  the condensate with the condensate is taken into account by
the terms with two $\psi$'s. Finally the self-interaction of the particles not  
in  the condensate
comes from the terms with no $\psi$ ($RR$ and $DD$).

We introduce the generalized potentials
\begin{eqnarray}
\label{32}
{\cal U}_{{\rm d}}({\bf x},{\bf y},t) &=& \delta({\bf x}-{
\bf y}) \int_{\bf z} \rho({\bf z},{\bf z},t) V({\bf x} -{\bf z
}) \nonumber \\
\label{31}
{\cal U}_{{\rm e}}({\bf x},{\bf y},t) & = & \rho ({\bf x},{
\bf y},t) V({\bf x}-{
\bf y}) \equiv {\cal U}_{{\rm e}}^{\rm r} + i {\cal U}_{{\rm e}}^{\rm i} \\
\label{33}
{\cal U}_{{\rm p}} ({\bf x},{\bf y},t) &=& \Delta({\bf x},{
\bf y},t) V({\bf x}-{\bf y}) \equiv {\cal U}_{{\rm p}}^{\rm r
} +i\; {\cal U}_{{\rm p}}^{\rm i} \nonumber 
\end{eqnarray}
\noindent where the notation emphasises real and imaginary parts 
of ${\cal U}_{\rm p}$. We also define the matrices
\begin{eqnarray}
\label{40}
A({\bf x},{\bf y},t) &=& h({\bf x},{\bf y}) -\mu + {\cal U}_{\rm p}^{\rm r}({\bf x},{\bf y},t)
 + {\cal U}_{\rm e}^{\rm r}({\bf x},{\bf y},t) + {\cal U}_{d}({\bf x},
{\bf y},t) \nonumber   \\
\label{41}
B({\bf x},{\bf y},t) &=& h({\bf x},{\bf y}) -\mu - {\cal U}_{\rm p}^{\rm r}({\bf x},{\bf y},t)
 + {\cal U}_{\rm e}^{\rm r}({\bf x},{\bf y},t) + {\cal U}_{d}({\bf x},
{\bf y},t)  \\
\label{42}
C({\bf x},{\bf y},t) &=& h({\bf x},{\bf y}) -\mu + {\cal U}_
{\rm e}({\bf x},{\bf y},
t) +{\cal U}_{d}({\bf x},{\bf y},t). \nonumber 
\end{eqnarray}
\noindent From Eqs. ({\ref 15}) and  (\ref{23}) we obtain 
 an abstract matrix form of the equations of motion
\begin{eqnarray}
\label{44}
\dot{\Sigma} &=& \frac{1}{8} G^{-1} A G^{-1} - 2 \Sigma A 
\Sigma - \frac{B}{2} +\{{\cal U}_{{\rm p}}^{\rm i},\Sigma \} - [{\cal U}_{\rm e}^{\rm i}, \Sigma]  \nonumber \\ 
\label{45}  
\dot{G} &=&  \{A,\{G,\Sigma\}\}  -  \{{\cal U}_{{\rm p}}^{
\rm i},G \} - [{\cal U}_{\rm e}^{\rm i}, G] \\ 
\label{46}
\imath \dot{\psi} &=& C \psi -  {\cal U}_{{\rm p}}  \psi^{
\ast}, \nonumber 
\end{eqnarray}
\noindent where
 we have used the fact that ($\Sigma$,G) are symmetric 
matrices. 
\noindent These equations (\ref{44})  are the nonlinear 
field equations for an
arbitrary interaction $V$ between the particles and
contain any external potential through  $h$. As an example 
the matrix product   $G^{-1} A G^{-1}$ can be written in 
coordinate representation as
\begin{equation}
\int_{{\bf z},{\bf w}}  G^{-1} ({\bf x},{\bf z},t) A({\bf z}
,{\bf w}) G^{-1} ({\bf w},{\bf y},t).
\end{equation}
The static equations  can be obtained by setting the canonical momenta to 
zero, that is $\Sigma({\bf x},{\bf t},t) = \pi({\bf x},t) =0$,
 $\dot{G}({\bf x},{\bf y},t) ={\dot{ \phi}}({\bf x},
t)=0$. From (\ref{22}) and  (\ref{29})  we than 
have
\begin{eqnarray}
\label{47}
R({\bf x},{\bf y},0)& \equiv & R({\bf x},{\bf y}) = \frac{1}
{2} \left[ \frac{G^{-1}({\bf x},{\bf y})}{
4} + G({\bf x},{\bf y}) - \delta({\bf x}-{\bf y}) \right] \nonumber \\
\label{48}
D({\bf x},{\bf y},0) & \equiv & D({\bf x},{\bf y}) =\frac{1}
{2} \left[ \frac{G^{-1}({\bf x},{\bf y})}{4} -G({\bf x},{\bf
 y})\right] \\
\label{48i}
\psi({\bf x},0) & \equiv & \psi({\bf x}) =\frac{\phi({\bf x}
)}{\sqrt{2}}. \nonumber 
\end{eqnarray}
\noindent
So that for the static case we have to self consistently solve
\begin{eqnarray}
\label{49}
&&\frac{1}{4} \int_{{\bf z},{\bf w}} G^{-1}({\bf x},{\bf z})
 A({\bf z},{\bf w}) G^{-1}({\bf w},{\bf y}) - B({\bf x},{\bf
 y}) = 0 \nonumber \\
\label{50}
\\
&& \int_{\bf z} \left[ B({\bf x},{\bf z}) \psi({\bf z}) - 2
\psi({\bf x}) \psi^{2}({\bf z}) V({\bf x}-{\bf z}) \right] =
 0. \nonumber 
\end{eqnarray}
\noindent using (\ref{27})-(\ref{40}). 
We note that if we constraint
$G=1/2$ for the static solution this leads to $R=D=0$ and $A=B$ so that equation (44)
is  to the usual non linear equation for the single quantity $\psi$ \cite{PT}  obtained from a
many body product wave function (permanent) for bosons.
However the time dependent eqs. (\ref{44}) are
more general, because our trial Gaussian is actually a coherent state
with an indefinite number of particles.

\section{Contact Interaction in the one dimensional uniform case}

\subsection{Gaussian Approximation}
We will specialize the results of the previous section for the one 
dimensional case and for a contact interaction so that
\begin{equation}
V(x-y) = \lambda \delta(x-y).
\end{equation}
\noindent Because of the existence of exact numerical solutions we will treat 
the uniform case so that the momentum representation is the natural choice
 where the   quantities $A$, $B$ and $G$ can simultaneously be
diagonalized and   
$\phi({k}) = \phi \delta({k})$. The static equations (\ref{49}) 
 become [we use the notation : $\int_{k} = \frac{1}{2 \pi} \int_{-\infty}^{\infty} dk$]
\begin{eqnarray}
\label{51}
\frac{1}{4} G^{-2}(k) A(k) -B(k)&=&0  \\
\label{52}
\phi \left[B(0) - \lambda \phi^{2} \right] &=&0.
\end{eqnarray}
\noindent
\noindent
  The static version of $A$ and $B$  from (\ref{40}) can be written as
\begin{eqnarray}
\label{53}
A({\bf k}) &=& e({\bf k}) - \mu + {\cal U}_{\rm p}  + 2  {\cal U}
 + \lambda \frac{\phi^{2}}{2} 
\nonumber  \\
&& \\
\label{54}
B({\bf k}) &=& e({\bf k}) - \mu - {\cal U}_{\rm p}  +2 {\cal U}
 + \frac{3}{2} \lambda  \phi^{2}. \nonumber  
\end{eqnarray}
\noindent where $e(k)=\frac{\hbar^{2} k^2}{2 m}$ and  the generalized 
potentials become 
\begin{eqnarray}
\label{55}
{\cal U}&\equiv& {\cal U}_{\rm d} = {\cal U}_{\rm e} = \lambda \int_{k'} R(k') \nonumber \\
\\
\label{56}
{\cal U}_{\rm p} &=& \lambda \int_{ k'} D(k') \nonumber 
\end{eqnarray}
\noindent So that we can write the solution for (\ref{51}) 
 as
\begin{equation}
\label{60}
G( k)= \frac{1}{2} \sqrt{\frac{A(k)}{B(k)}}  
\end{equation}
\noindent and for (\ref{52}) we have
\begin{eqnarray}
\label{61aa}
&&\phi=0 \\
&& {\rm or} \nonumber \\
\label{61aaa}
&& B(0) =  \lambda \phi^{2}
\end{eqnarray}
\noindent using (\ref{60}) with  (\ref{47})  we can express $D$ 
and $R$ as functions of $A$, $B$ 
\begin{eqnarray}
\label{61a}
D(k) &=& \frac{1}{2} \left[\frac{G^{-1}(k)}{4} - G(k)  \right] = \frac{1}{4} \frac{ B(k) - A(k) }{\sqrt{A(k) B(
k)}} \nonumber  \\
\label{62}
\\
R(k) &=& \frac{1}{2} \left[\frac{G^{-1}(k)}{4} + G(k) - 1 \right] =  \frac{1}{2} \left\{ \frac{B(k)+A( k) }{2 \sqrt{A(
k) B(k)}} -1 \right\}. \nonumber 
\end{eqnarray}
\noindent  From (\ref{53}),(\ref{54}) and (\ref{60}) we see that 
$A(0)$ and $B(0)$ must be positive so that if $\lambda <0$ equation ({\ref{61aaa}) 
demands that $B(0) <0$ which is inconsistent with the previous 
statement. So the only possible solution in this case is $\phi=0$. For 
$\lambda >0$ the symmetry breaking solution $\phi \ne 0$ , using (\ref{61aaa}), gives us  
\begin{equation}
\label{62b}
\mu = \lambda \int_{k'} \left[ 2 R(k') - D(k') \right] + \frac{\lambda}{2} \phi^{2}.
\end{equation}
\noindent Having solved for $\mu$  we can rewrite $A$ and $B$ in (\ref{53}) and (\ref{54}) as
\begin{eqnarray}
\label{62bb}
A(k) &=& e(k) + 2 \lambda \int_{k'} D_{k'} \equiv  e(k) +
2 \lambda a \nonumber \\
\\
\label{62b1}
B(k) &=& e(k) + \lambda \phi^{2} \equiv  e(k) + 2 \lambda b. \nonumber 
\end{eqnarray}
\noindent On the other hand using (\ref{61a}) and (\ref{62bb}) we can write 
a pair  of non-linear equations for $a$ and $b$  
\begin{eqnarray}
\label{62c}
a & =& \frac{\lambda}{2} \int_{{k}'} \frac{\left[b-a \right]}{\sqrt{
\left[e(k') + 2 \lambda a\right] \left[e(k') + 2 \lambda b \right]}} \nonumber \\
\label{62d}
\\
b& =&   \rho - \frac{1}{2}  \int_{{k}'} \left\{\frac{e(k) +\lambda \left[b +
a \right]} { \sqrt{\left[
e(k') + 2 \lambda a\right] \left[e(k') + 2 \lambda b \right]}}-1 \right\}, \nonumber
\end{eqnarray}
\noindent where  we have used the total density constraint 
\begin{equation}
\rho =  \frac{\phi^{2}}{2} + \int_{ k'} R(k').
\end{equation}
\noindent which actually becomes our scale. 
This non-linear set of equations can be solved for a given $\rho$,
given  $a$ and $b$. Once we have calculated $a$ and $b$  we can compute the
chemical potential through  
\begin{equation}
\mu =  \lambda \left[2 \rho - a -  b \right]. 
\end{equation}
\noindent In the same fashion the ground state energy  density ($E/L = ({\cal H})/L $)  can be computed
obtaining
\begin{equation}
\label{62e}
E/L = \frac{\lambda}{2} a^{2} - \frac{\lambda}{2} b^{2} - \lambda a b + \lambda \rho^{2} + K,
\end{equation}
\noindent where $K$, the contribution from the kinetic energy can also be computed in terms of $a$ and $b$ as
\begin{equation}
\label{62f}
K = \frac{1}{2} \int_{k'} e(k') \left\{\frac{e(k') + \lambda[a+b]}{\sqrt{\left[e(k') + 2 \lambda a\right] 
\left[e(k') + 2 \lambda b \right]}}-1 \right\}
\end{equation}

As an aside we remark that for a  dilute system we can approximate the self-consistent 
equations for $a$ and $b$
by truncating them at a second iteration. A first iteration on (\ref{62c})
takes $a \approx 0$,
$b \approx \rho$ which implies  zero depletion and ${\cal U}_{\rm p} = 0$. This  leads us to
a non-pairing theory (Gross Pitaevskii equation \cite{PT}). Then the next
iteration leads to
\begin{eqnarray}
a &\approx& \frac{\lambda \rho}{2}  \int_{k'} \frac{1}{\sqrt{e(k')^{2} + 2
\lambda \rho e(k')}} \nonumber \\
\\
b &\approx&  \rho - \frac{1}{2}  \int_{k'} \left\{\frac{e(k') +
\lambda \rho}{\sqrt{e(k')^{2} + 2
\lambda \rho e(k')}}
-1 \right\}.  \nonumber 
\end{eqnarray}
\noindent Then we can   calculate $R$  and $D$ truncating the self-consistency and giving  the same
results as the Bogoliubov scheme. Physically this  means neglecting
the effect of the terms that take into account
the self-intercation of the particles not in  the condensate ($DD$ and $RR$).This approximation
 is usually valid for dilute systems where these terms are not important.
With this truncation the ground state energy can be easily computed giving
\begin{equation}
\label{b11}
\frac{E}{N} = \frac{\lambda}{2}\rho \left[1 - \frac{4}{3 \pi} \sqrt{\gamma}
\right],
\end{equation}
\noindent where the dimensionless parameter $\gamma$ is
\begin{equation}
\label{b2}
\gamma = \frac{\lambda m}{\rho \hbar^{2}}.
\end{equation}

Returnig to our discussion we determine the excitations through the RPA equations which can be found 
by expanding all
quantities
around their equilibrium value \cite{FN}. Thus we write

\begin{eqnarray}
\label{64}
G({k},{k}',t)  &=&  G({k})  \delta({k}-{
k}')  + \delta G({k},{ k}',t) \nonumber  \\
\label{65}
\Sigma({k},{k}',t) &\rightarrow& \delta{\Sigma}({k},{ k
}',t)   \\
\label{66}
\phi({k},t) &=& \phi \delta({k}) + \delta \phi({
k},t)  \nonumber \\
\label{67}
\pi ({k},t) & \rightarrow & \delta \pi ({k},t). \nonumber 
\end{eqnarray}
 \noindent Thus we have  written $G$ and $\Sigma$ in the 
basis where the equilibrium $G$ is diagonal and kept terms up
 to first order in small quantities, of course  
 the diagonal basis is  plane waves.

It will be useful to introduce new momentum coordinates so 
that
\begin{eqnarray}
\label{68}
{P} &=& {k} - {k}' \\
\label{69}
{q} &=& \frac{{k}+{k}'}{2}
\end{eqnarray}
\noindent  and
 \begin{equation}
\label{70}
\delta G({k},{k}') \rightarrow \delta G({P},{
 q}).
\end{equation}
\noindent We will see that  ${ P}$ and ${ q}$ can be 
interpreted as total and relative momenta respectively 
of a pair of quasi-bosons.  
 We can then write the RPA equations in a form where ${ P}$
 is diagonal and can be considered as a dummy variable
\begin{eqnarray}
\label{71}
\delta \dot{G}({q},{ P},t) &=&\! \! \!s_{\rm K}({
q},{P}) \delta \Sigma({ q},{P},t) +  c_{\rm K} ({
 q},{P}) \delta \pi({P},t) + \int_{{q}'}S_{
\rm K}({q},{q}',{P}) \delta \Sigma({q}',{ P
},t) \nonumber  \\
\label{72}
-\delta \dot{\Sigma}({q},{P},t) &=& \! \! \!s_{\rm M
}({q},{P}) \delta G({q},{P},t) + c_{\rm M}({
q},{P}) \delta \phi({ P},t) + \int_{{q}'}S_{
\rm M}({q},{q}',{P}) \delta G({q}',{P},t)
  \\
\label{73}
\delta \dot{\phi}({ P},t) &=& \! \! \!   \delta \pi({
P},t) A({P}) + \int_{{q}'} c_{\rm K}({q}',{P
}) \delta \Sigma({q}',{P},t) \nonumber \\
\label{74}
-\delta \dot{\pi}({ P},t) &=& \! \! \!
\delta \phi({P},t) B({P}) + \int_{{q}'} c_{\rm M
}({q}',{P}) \delta G({q}',{ P},t). \nonumber 
\end{eqnarray}
\noindent We note that for a given value of $P$  
the $(\pi,\phi)$ degree of freedom
 is coupled to the much more numerous degrees of freedom
 $(\Sigma,G)$ which are labeled by ${q}$. Different ${
q}$ values among $(\Sigma,G)$ are also coupled.
\noindent Introducing the notation
$f({q}' + {P}/2) = f'_{+}$ and $f({ q} - { P}/
2) = f_{-}$,  we find non-diagonal matrices in $({ q},{
 q}')$
\begin{eqnarray}
\label{75}
S_{\rm K}({\bf q},{\bf q}',{\bf P})& = &\lambda \left[G_{+} +
G_{-} \right] 
 \left[G'_{+} +
G'_{-} \right] + \lambda \left[G_{+} - G_{-} \right] \left[G'_{+} - G'_{-} \right] \nonumber  \\
\label{78}
\\
S_{\rm{M}}({\bf q},{\bf q}',{\bf P}) & = &  \frac{\lambda}{2}+
\frac{\lambda}{4} \left[1 - \frac{G^{-1}_{+}
G^{-1}_{-}}{4} \right] 
\left[1 - \frac{G^{'^{-1}}_{+}
G^{'^{-1}}_{1}}{4} \right] \nonumber \\
&& + \left[ \frac{G^{-1}_{+}
G^{-1}_{-}}{4} \right]
 \left[ \frac{G^{'^{-1}}_{+}G^{
'^{-1}}_{-}}{4}
\right] \nonumber 
\end{eqnarray}
\noindent and diagonal elements
\begin{eqnarray}
\label{76}
s_{\rm K}({\bf q},{\bf P})& = & 2 \left[ A_{+}
G_{-} +  A_{-}
G_{+} \right] \nonumber  \\
\\
\label{79}
s_{\rm M}({\bf q},{\bf P})& = & \frac{G^{-2}_{+} G^{-1}_{-}
A_{+} + G^{-2}_{-} G^{-
1}_{+}
A_{-}}{8}. \nonumber
\end{eqnarray}
\noindent Finally we see the coupling elements between $(\pi
,\phi)$ and $(\Sigma,G)$
\begin{eqnarray}
\label{77}
c_{\rm K}({\bf q},{\bf P})& = & \lambda \phi \left[G_{+} +
G_{-} \right] \nonumber \\ 
\\
c_{\rm M}({\bf q},{\bf P})& = & \frac{\lambda}{2} \phi \left[3 -   
 \frac{
G^{-1}_{+}
G^{-1}_{-}}{4} \right], \nonumber 
\end{eqnarray}
\noindent which vanish when the symmetry in $\phi$  is conserved $(\phi
=0)$. As pointed out above the equations are diagonal in ${
\bf P}$ so we can  interpret it as the total momentum of a 
pair of quasi-bosons. Because $\delta G$,  $\delta \Sigma$ and
 $\delta \phi$,  $\delta \pi$ are canonical variables we may
 invert the definitions of  momentum and coordinate.
For convenience, we define column vectors
\begin{equation}
\label{81}
\Theta({q},{P},t) = \left( \begin{array}{c}
                                    \delta \Sigma({q},{
 P},t) \\
                                    \delta \pi({P},t)
                                   \end{array}
                              \right),
\; \; \; \; \Pi({q},{P},t) =-\left( \begin{array}{c}
                                        \delta G({q},{
P},t) \\
                                        \delta \phi({P},
t)
                                       \end{array}
                                \right).
\end{equation}
\noindent Then  we can write a coupled oscillator Hamiltonian that
 corresponds to the RPA equations of motion in a suggestive matrix 
element form
\begin{equation}
\label{82}
H_{\rm RPA} = \frac{1}{2} \widetilde{\Pi} M^{-1} \Pi + \frac{1}{2} \widetilde{\Theta} K \Theta  
\end{equation}
\noindent where the matrixes $M^{-1}$ and $K$ are the 
generalizations of oscillator mass and spring constant
\begin{equation}
\label{83}
K =  \left( \begin{array}{cc}
              S_{\rm K} + s_{\rm K} & c_{\rm K} \\
              c_{\rm K}         & A
             \end{array}
     \right),
\; \; \; \; M^{-1} = \left(\begin{array}{cc}
              S_{\rm M} + s_{\rm M} & c_{\rm M} \\
              c_{\rm M}         & B
                      \end{array}
      \right).
\end{equation}
\noindent We may separate  the diagonal part of $H_{\rm RPA}$ so 
that
\begin{equation}
\label{84}
H_{\rm RPA} = H_{0} + H_{\rm int},
\end{equation}
\noindent where
\begin{eqnarray}
\label{85}
H_{0} &=& \frac{1}{2}  \left(\matrix{\delta \Sigma^{\ast}&\delta \pi^{\ast}}
 \right)\left(\matrix{s_{\rm K}&0\cr 0&A} \right)\left(
\matrix{\delta\Sigma\cr \delta \pi}\right)+\frac{1}{2} \left(
\matrix{\delta G^{\ast}&\delta \phi^{\ast}} \right)\left(\matrix{s_{\rm M}&0
\cr 0&B} \right)\left(\matrix{\delta G\cr \delta \phi}\right) \nonumber  \\
&&  \\ 
H_{\rm int} &=& \frac{1}{2}  \left(\matrix{\delta \Sigma^{\ast}&
\delta \pi^{\ast}} \right)\left(\matrix{S_{\rm K}&c_{\rm M}\cr c_{\rm M}
&0} \right)\left(\matrix{\delta\Sigma\cr \delta \pi}\right)+
\frac{1}{2} \left(\matrix{\delta G^{\ast}&\delta \phi^{\ast}} \right)\left
(\matrix{S_{\rm M}&c_{\rm M}\cr c_{\rm M}&0}  \right)\left(
\matrix{\delta G\cr \delta \phi}\right). \nonumber 
\end{eqnarray}
\noindent Introducing the trivial canonical 
transformation
\begin{eqnarray}
\label{89}
\delta \Sigma &\rightarrow &  \sqrt{s_{\rm M}}\;\; \delta 
\Sigma \;\;\;\;\;\;
\delta G  \;\rightarrow\;  \frac{\delta G}{\sqrt{s_{\rm M}}}
\nonumber\\
&& \\
\label{90}
\delta \pi &\rightarrow&  \sqrt{B}\;\; \delta \pi \;\;\;\;\;
\;
\delta \phi \; \rightarrow \; \frac{\delta \phi}{\sqrt{B}} \nonumber
\end{eqnarray}
\noindent we obtain a simpler form for the diagonal part
\begin{equation}
\label{91}
H_{0} = \frac{1}{2}  \left(\matrix{\delta \Sigma^{\ast} &\delta \pi^{\ast}
} \right)\left(
\matrix{s_{\rm M} s_{\rm K}&0\cr 0&A B } 
\right)\left(\matrix{\delta\Sigma \cr \delta \pi }\right)+
\frac{1}{2} \left(\matrix{\delta G^{\ast} &\delta \phi^{\ast}} \right)\left
(\matrix{1&
0\cr 0&1} \right)\left(\matrix{\delta G\cr \delta \phi}
\right).
\end{equation}
\noindent If we define  $\Omega_{1}$ and $\Omega_{2}$ 
\begin{eqnarray}
\label{93}
\Omega_{1}({\bf P}) & =& \sqrt{A({P}) B({P})}
\\
\label{94i}
\Omega_{2}({q},{P}) &=& \sqrt{s_{\rm K}({q},{
P}) s_{\rm M}({q},{P})}.
\end{eqnarray}
\noindent and  use the definitions of $s_{\rm K}$ and $s_{
\rm M}$ from (\ref{76})  we get, after some 
algebra, the remarkable result
\begin{equation}
\label{94}
\Omega_{2}({q},{P}) = \sqrt{A_{+} B_{+}} + 
\sqrt{A_{-} B_{-}}= \Omega_{1}({k}) + \Omega_{1}({k
}'),
\end{equation}
\noindent so that $\Omega_{1}({P})$ and $\Omega_{2}({q},{P})$
 can be interpreted as
the one and two free quasi-boson energies. We note that
$\Omega_{2}(0,{P}) = 2 \Omega_{1}({P}/2)
$, which means that at zero relative momentum $\Omega_{2}({P},0)$
 corresponds to two quasi-bosons  with momentum
 ${P}/2$. Thus 
the oscillations of the $\delta \phi$, $\delta \pi$
pair can be interpreted as a  quasi-boson mode while
 the oscillations
of $\delta G$, $\delta \Sigma$ can be interpreted as an 
interacting pair of
these same quasi-bosons. When $\phi=0$, we get $c_{\rm K}=c_{\rm M} 
 = 0$ and the one
and two quasi-bosons systems are calculated independently. 
When $\phi \ne 0$
we must rediagonalize so that our final modes will be 
mixtures of one and
two quasi-bosons.
The variable $ {q}$ represents the internal motion of 
the quasi-boson
pair
 with interaction given by the quantities $S$. In general 
this is a
scattering
problem and we must search for the scattering amplitude at a
 given energy  and
 $P$, where the asymptotic  conditions are determined by
(\ref{94i}). In addition the coupling  of one and two quasi-bosons 
will always lead to a bound state which is a particular mixture 
of the one quasi-boson mode with a bound component of the two quasi-bosons. 

\noindent As we did for the static results it is straightforward to see that the
truncation that gives the Bogoliubov results implies  neglecting the
coupling ($H_{\rm int} =0$) and will lead to the usual result
\begin{equation}
\label{bogexc}
\Omega_{1}^{\rm b}(P) = \sqrt{e^{2}(P) + 2 \lambda \rho e(P)}.
\end{equation}
\noindent We can see that the Bogoliubov excitations consider no interaction
between
the quasi-bosons. Note that because the Bogoliubov does not take into account
the
self-interaction of the particles not in  the condensate we have   $k'=0$ which means
$P=k$.

Returning to the dispersion relation for the bound mode one can finally eliminate 
the two quasi-boson components. To see this we try oscillatory solutions for  
(\ref{71})  such as
\begin{eqnarray}
\Theta(t) &=& \Theta e^{i \Omega t} \nonumber \\
&& \\
\Pi(t) &=& \Pi  e^{i \Omega t}, \nonumber 
\end{eqnarray}
\noindent and equations (\ref{71}) can be written in a compact form  
\begin{equation}
\label{s1}
{\cal M}.{\cal X} = {\cal Y},
\end{equation}
\noindent that is 
\begin{equation}
\left(
\begin{array}{cccc}
-\Omega & -s_{\rm K} & 0 & -\lambda x \phi \\
s_{\rm M}  & \Omega & \lambda v \phi & 0  \\
0      & 0      & - \Omega & -A \\
0      & 0 & B &  \Omega
\end{array} \right).
\left(
\begin{array}{c}
\delta G \\
\delta \Sigma \\
 \delta \phi \\
\delta \pi
\end{array} \right)
=
\left(\begin{array}{c}
\lambda (x X + r R)    \\
-\lambda y  Y - 2 \lambda z Z    \\
\lambda \phi X    \\
- \lambda \phi (Y + 2 Z)
\end{array} \right)
\end{equation}
\noindent where for simplicity we have used
\begin{eqnarray}
r(q,P) &=& G_{+} -  G_{-} \nonumber \\
x(q, P) &=& G_{+} + G_{-} \nonumber \\
y(q, P) &=& \frac{1}{2} \left[1 +  \frac{G_{+}^{-1} G_{-}^{-1}}
{4} \right] \\
z(q,P) &=& \frac{1}{2} \left[1 -  \frac{G_{+}^{-1} G_{-}^{-1}}
{4} \right] \nonumber \\
v(q,P) &=& y(q,P) + 2 z(q,P) \nonumber  .
\end{eqnarray}
and also
\begin{eqnarray}
\label{s2}
R(P) &=&  \int_{q'} r( q', P)  \delta \Sigma(q',P) \nonumber \\
\label{s22}
X(P) &=&   \int_{q'}  x( q', P)  \delta \Sigma(q',P)  \\
\label{s3}  
Y(P) &=&   \int_{q'}  y( q', P)  \delta G(q',P) \nonumber \\
\label{s4}
Z(P) &=&   \int_{q'}  z( q', P)  \delta G(q',P) \nonumber
\end{eqnarray}
\noindent Note that the condition $det{\cal  M} =0$ from the 
homogeneous equation (${\cal Y}=0$) gives us back
 $\Omega_{1}$ and $\Omega_{2}$  discussed above. 

In the discussion which follows we look for the bound state 
referred to above by holding $\Omega < \Omega_{1} < \Omega_{2} $ 
so that it is not necessary here to include the usual scattering 
$i \epsilon$ in the denominators.
\noindent We can invert 
${\cal M}$ obtaining 
\begin{eqnarray}
\label{s5}
\delta G &=& -\frac{\lambda  \Omega r R}{(\Omega^{2} - \Omega_{2}^{2})} -\left[\frac{x}{(\Omega^{2} - \Omega_{2}^{2})} +  \lambda \phi^{2}
  \frac{ s_{\rm K} v  + B x}{(\Omega^{2}-\Omega_{1}^{2})(\Omega^{2} - \Omega_{2}^{2})}\right
] \lambda \Omega X \nonumber \\
&& + \lambda^{2} \phi^{2} \frac{ A s_{\rm K} v +x \Omega^{2}}{(\Omega^{2}-\Omega_{1}^{2}
)(\Omega^{2} -\Omega _{2}^{2})} (Y + 2 Z) +  \lambda s_{\rm K}  \frac{y Y + 2 z Z}{
(\Omega^{2} -\Omega _{2}^{2})} \nonumber \\
 \nonumber \\
\label{s6}
\delta \Sigma &=& \frac{\lambda s_{\rm K} r R}{(\Omega^{2} - \Omega_{2}^{2})} + \left[\frac{ s_{\rm M} x }{\Omega^{2} -\Omega _{2}^{2}} + \lambda 
\phi^{2} \frac{ B  s_{\rm M} x + v \Omega^{2}}{(\Omega^{2}-\Omega_{1}^{2})(\Omega^{2} - 
\Omega_{2}^{2})}\right] \lambda X \nonumber \\
&& - \lambda^{2} \phi^{2} \Omega \frac{A v +  s_{\rm M} x }{(\Omega^{2} -\Omega _{2}^{2})(\Omega^{2} - \Omega _{1}^{2})}(Y + 2 Z)
 - \lambda \Omega \frac{y Y + 2 z Z}{\Omega^{2} -\Omega _{2}^{2}} \nonumber \\
\\
\label{s7}
\delta \phi &=& -\lambda \phi \frac{\Omega X - A(Y + 2 Z)}{(\Omega^{2}-\Omega_{1}^{2})}
\nonumber \\
\label{s8}
\delta \pi &=& -\lambda \phi \frac{\Omega (Y + 2 Z) - B X}{(\Omega^{2}-\Omega_{1}^{2})} \nonumber 
\end{eqnarray}
\noindent Now we substitute (\ref{s5} ) in the 
definitions of the quantities $R$, $X$, $Y$  and $Z$ (\ref{s2})
. Because $r(q,P)$ is an odd function i.e $r(-q,P) = - r(q,P)$ it is 
easy to check  we end up with a   linear and homogeneous 
system for $X$, $Y$ and $Z$
, that looks like
\begin{equation}
\label{s10}
W({P},\Omega ). F = 0
\end{equation}
\noindent where, omiting the ${P}$ dependence we have
\begin{equation}
\label{s11}
W =
\left(
\begin{array}{ccc}
W_{1,1} & W_{1,2} & W_{1,3} \\
W_{2,1} & W_{2,2} & W_{2,3} \\
W_{3,1} & W_{3,2} & W_{3,3}
\end{array} \right),
\; \; \; \; \; \; \; \;\;\;
F =
\left(
\begin{array}{c}
X \\
Y \\
Z
\end{array} \right)
\end{equation}
\noindent where the elements of $W$ are given in the  appendix B.
The system (\ref{s10}) will have a non-trivial solution if 
\begin{equation}
\label{s111}
{\rm det} W(P,\Omega (P)) =0
\end{equation}
so that we have  for each $ P$ the corresponding energy $\Omega(P)$.  
Numerically the problem reduces to calculating determinants 
of a $3 \times 3$ matrix.

A very general property \cite{KP} of the dispersion 
relation $\Omega(P)$ can be 
proven for the particular case where $P = 0$. 
In this case  the first line of the matrix (\ref{s11}) 
using (\ref{62b1}) and (\ref{94i}) is 
\begin{equation}
W_{1,1} = 1 - \lambda \frac{A(0)-B(0)}{A(0)} \int_{q} \frac{x^{2}({ q},0
)}{s_{\rm K}({ q},0)}, \;\;\;\; 
W_{1,2} = 0, \;\;\;\; 
W_{1,3} = 0
\end{equation}
\noindent and now using (\ref{62bb})-(\ref{62c})   we have that
\begin{equation}
 \lambda \int_{\bf q} \frac{x^{2}({ q},0)}{s_{\rm K}(q,0)} = \frac{\lambda}{2} 
\int_{q}  \frac{1}{\sqrt{A({ q}) B({q})}} = \frac{A(0)}{A(0)-B(0)}.
 \end{equation}
\noindent So that the first line of the matrix is zero, making the determinant
 vanish, showing that we will always have a gapless dispersion relation 

independently of the value of $\lambda$,
\begin{equation}
\Omega(0) =0.
\end{equation}
\noindent This zero mode of the RPA equations is the standard Goldstone 
mode as it's structure is 
associated with the symmetry breaking (indefinite particle number) by the 
trial wave functional \cite{KP}.

We note that the dispersion relation that comes from the RPA depends on 
the total momentum defined in (\ref{68}) which means that $\Omega$ is 
function of $(k-k')$, so that it takes into account that to obtain 
an excitation we remove a particle with momentum $k'$ and create a 
particle with momentum $k$. This result is very different from the 
Bogoliubov excitation where $k'$ is taken to be zero. In other words 
the Gaussian approximation takes into account the effect of the 
depletion on the excitations.

\subsection{The Exact Solution}
All the results summarized in this section were obtained by Lieb 
\cite{LB1}-\cite{LB2} who calculated the exact ground state energy 
and also the excitations in terms of independent particle and hole excitations. Our purpose 
is to make a connection with this work which showed that interacting bosons 
in one dimension  can be analogous to a fermi gas. We will show how our 
modes correspond to Lieb's particle-hole excitations.

First of all when $\lambda \rightarrow \infty$ it is possible to 
recover the well-known result \cite{GR1} that 
for an infinite coupling constant interacting bosons behave 
like a system of free fermions, so that the ground state energy can be easily computed by
\begin{equation}
\frac{E}{N} =  \frac{1}{\rho} \frac{1}{2 \pi} \int_{-K_{\rm f}}^{K_{\rm f}}  e(k) dk 
\end{equation}
\noindent where $K_{\rm f}$ is the fermi- momentum. In this case it is trivial to calculate
\begin{equation}
\rho = \frac{1}{2 \pi} \int_{-K_{\rm f}}^{K_{\rm f}} dk   
\end{equation}
\noindent which gives $K_{\rm f} = \pi \rho$. So that in this particular limit
\begin{equation}
\frac{E}{N} =  \frac{\hbar^{2}}{2 m} \frac{\pi^{2} \rho^{2}}{3}
\end{equation}

We can divide the particle hole excitations of the free fermi gas into
two parts. The particular hole
excitations
that
correspond to removing a particle from an occupied state to just above the fermi
level ($K_{\rm f}$)
\begin{equation}
\epsilon_{\rm h}(k') = \frac{\hbar^{2} \pi^{2} \rho^{2}}{2 m} -
\frac{\hbar^{2} k'^{2}}{2 m}
\end{equation}
\noindent where $K_{\rm f} <k' < K_{\rm f}$ and the particle
excitations where
we remove a
particle from the fermi level to an unoccupied state
\begin{equation}
\epsilon_{\rm p}(k) = \frac{\hbar^{2} k^{2}}{2 m} -
\frac{\hbar^{2} \pi^{2} \rho^{2}}{2 m}
\end{equation}
\noindent where $k>K_{\rm f}$ or $k<-K_{\rm f}$. To
produce a particle hole
excitation we must add
these excitations giving us
\begin{equation}
\epsilon_{\rm p}(k,k') = \frac{\hbar^{2} k^{2}}{2 m} -
\frac{\hbar^{2} k'^{2}}{2 m}
\end{equation}
Using  $\epsilon_{\rm h}$ and $\epsilon_{\rm p}$ one
can look at $E(k',k)$ as a
function
of the total and relative momentum  $P=k-k'$ and $q=(k+k')/2$.
Because the hole excitations are limited to
($-K < k' < K $) we need to separate two cases, for $P>0$ we have
\begin{eqnarray}
\mbox{if $P < 2 K_{\rm f}$} &\rightarrow&
\left\{ \begin{array}{l}
                                          k = K_{\rm f}     \\
                                         k' = K_{\rm f} - P
                                         \end{array} \right. \nonumber \\
&& \\
\mbox{if $P>2 K_{\rm f}$} &\rightarrow&
\left\{ \begin{array}{l}
                                         k = P - K_{\rm f}  \\
                                            k' = -K_{\rm f}
                                   \end{array} \right. \nonumber
\end{eqnarray}
\noindent The first case means fixing the
momentum of the particle and moving
the momentum of the hole
while in the second one we fix the
momentum of the hole at it lowest possible
value and
move the momentum of the particle. So that once we
know $\epsilon_{\rm h}$ and
$\epsilon_{\rm p}$ the threshold curve defined by the
lowest value of $E(k',k)$ for a given $P$ wil
 be  given by
\begin{equation}
\label{ee1}
E(P) = \left\{ \begin{array}{ll}
           \epsilon_{\rm h}(K_{\rm f} - P) +
\epsilon_{\rm p}(K_{\rm f}) &  \mbox{for $P<  2 K_{\rm f}$}  \\
             \epsilon_{\rm h}(-K_{\rm f}) +
\epsilon_{\rm p}( P - K_{\rm f}) &  \mbox{for $P> 2 K_{\rm f}$}.
               \end{array} \right.
\end{equation}
\noindent Which gives us
\begin{equation}
\label{ee2}
E(P) = \left\{ \begin{array}{ll}
               2 \pi \rho P - P^{2} &    \mbox{for $P<  2 K_{\rm f}$}  \\
               P^2 + 2 \pi \rho P &  \mbox{for $P> 2 K_{\rm f}$}
               \end{array} \right.
\end{equation}
\noindent Note that the $E(P)$ curve contain two parts. Finite range part $P<  2 K_{\rm f}$
where the contribution comes basically  from the hole excitations.
Infinite range part $P> 2 K_{\rm f}$ from the particle excitations. The more dilute the system
the less finite range part in $E(P)$.

The generalization for bosons interacting with a finite $\lambda$ was
carried out by Lieb and
Lininger\cite{LB1}, \cite{LB2}. The showed the ground-state can be calculated as
\begin{equation}
\label{l1}
\frac{E}{N} = \frac{1}{\rho} \int_{-K}^{K} f(k) e(k) dk
\end{equation}
\noindent where $f(k)$ is the solution of
\begin{equation}
\label{l11}
2 \gamma \rho  \int_{-K}^{K} \frac{f(p)}{\rho^{2} \gamma^{2} + (p-k)^{2}} dp = 2 \pi f(k) -1
\end{equation}
\noindent with $\gamma$  given in (\ref{b2}). The condition
\begin{equation}
\label{l2}
\int_{-K}^{K} f(k) dk = \rho
\end{equation}
\noindent determines $K$. For the excitations Lieb defined two different basic interactions which  he
called "particle" and "hole" excitations. To determine these excitation Lieb
showed that it was sufficient to solve new
integral equations. For the particle energy
\begin{equation}
\label{exc1}
\epsilon_{\rm p}(k) = \frac{\hbar^{2} k^{2}}{2 m} - \mu + \frac{\hbar^{2}}{m} \int_{-K}^{K} p
J_{\rm p}(p) dp
\end{equation}
\noindent where $J_{\rm p}(p)$ could  be obtained by solving
\begin{equation}
2 \pi J_{\rm p}(p) = 2 \gamma \rho \int_{-K}^{K} \frac{J_{\rm p}(r)}{\rho^{2} \gamma^{2} + (p-r)^{2
}} dr - \pi + 2 \tan^{-1}\left[\frac{
k-p}{\gamma \rho} \right]
\end{equation}
\noindent  and for the hole energy $\epsilon_{\rm h}(P,\lambda)$ he had
\begin{equation}
\epsilon_{\rm h}(k') = \mu -  \frac{\hbar^{2} k'^{2}}{2 m} + \frac{\hbar^{2}}{m} \int_{-K}^{K} p J_
{\rm h}(p) dp.
\end{equation}
\noindent where  $J_{\rm h}(p)$ was obtained from
\begin{equation}
\label{exc2}
2 \pi J_{\rm h}(p) = 2 \gamma \rho \int_{-K}^{K} \frac{J_{\rm h}(r)}{\rho^{2} \gamma^{2} + (p-r)^{2
}} dr + \pi - 2 \tan^
{-1}\left[\frac{k'-p}{\gamma \rho} \right].
\end{equation}
\noindent In the limit $\lambda \rightarrow \infty$ we can see that  $J(p) \rightarrow
0$ and $\mu
\rightarrow (\hbar^{2} K^{2})/(2 m)= (\hbar^{2} \pi^{2} \rho)/(2 m)$ recovering the free fermion
results.
Looking at the expressions for the ground state energy and the excitations for
a given
$\lambda$,  Lieb  interpreted  them as those of a quasi-fermi gas  where $K$ is an
interaction dependent
fermi momentum  and the distributions factors $f(k)$ and $J(k)$ given a special weight for each $k$.
Using this analogy is very reasonable, since it is correct in both the $\lambda=0$ and $\lambda = \infty$.
To obtain the threshold curve $E(P)$ we can use (\ref{ee1}). The difference is that
now $\epsilon_{\rm p}$ and $\epsilon_{\rm h}$
 will have different curvatures depending on the interaction.
Note that in his original work Lieb compared the Bogoliubov scheme with particle
and holes excitations
separately and got very good agreement with the particle  excitations. This
is interesting,
 as we pointed out earlier, and just tells us that
ithe Bogoliubov scheme does not contain hole excitations.  In general as $\gamma$ increases the
contribution of the holes  for the particle-hole excitation energy becomes more and more
important and this effect is in part described by the Gaussian theory.

\section{Numerical Results and Conclusions}

For the numerical computations we follow Lieb and use the 
dimensionless coupling constant $\gamma$ and scale all 
lengths by $\rho$ and all energies by $(\hbar^{2} \rho^{2})/(2 m)$.
In these  units we can write the ground state energy per particle as
\begin{equation}
\frac{E}{N} =  g(\gamma).
\end{equation}
\noindent For the Gaussian static results we solve the  
non-linear system (\ref{62c})-(\ref{62d}) and determine $a$
 and $b$ for
$0 < \gamma <10$.  After computing $K$ defined in (\ref{62f}) 
one can get $g(\gamma)$ using (\ref{62e}). In the 
Bgoliubov scheme (\ref{b11}) leads to
\begin{equation}
g_{\rm B}(\gamma) = \gamma \left[1 - \frac{4}{3\pi} \sqrt{\gamma} \right]
\end{equation}
\noindent Finally the exact result for $g(\gamma)$ was obtained by Lieb 
solving (\ref{l1})-(\ref{l2}). The results
 as a function of $\gamma$ can be seen in Fig. 1. Where 
one can basically see the very good agreement of both 
Gaussian and Bogoliubov) for low $\gamma$ ($\gamma \leq 1$) with the 
difference that by construction  the Gaussian result
lt is always an upper bound and the Bogoliubov energy is below 
the exact result. For higher values of $\gamma$ the
Bogoliubov results, as expected, collapses ($\gamma \sim 5$)  
while the Gaussian theory still gives a result. In the
 hard core limit ($\gamma \rightarrow \infty$) the 
Gaussian approximation fails to go to the  finite free fermi 
energy of the exact solution.

To obtain the "exact" excitation energies  we used  
Lieb's results for $\epsilon_{\rm h}$ and $\epsilon_{\rm p}$ 
obtained by solving (\ref{exc1})-(\ref{exc2}). Using (\ref{ee1}) 
we obtain the excitation curves $E(P)$ as a function
 of $P$ as shown in Fig. 2 for $\gamma = 0,0.787,4.527$ 
and $\infty$. Note that for $P < 2 K$ the contribution from
 the hole excitations produce quite interesting dispersion curves.

To illustrate the RPA results we plot in Fig.3,  $\Omega_{1}$, $\Omega_{2}$ 
and $\Omega$ as a function of $P$ for $\gamma = 7.551$. The 
relative momentum  variable $q$ will gives us a continuum that corresponds 
to a scattering region and one can see clearly how $\Omega_{1}$ 
gets pushed down by the interactionp producing the gapless $\Omega$.  
We compare the bound state dispersion relation $\Omega$  with the 
Bogoliubov one and with the exact results for  $\gamma=0.787 $ (Fig. 4) 
and $\gamma = 4.527$ (Fig. 5). 
We note that the improvement of the RPA compared to the 
Bogoliubov result increases with $\gamma$. When $\gamma$ goes 
to infinity both the $RPA$ and the Bogoliubov schemes fail.

The results of this paper can be summarized in three points

1) We have seen that the Gaussian variational method takes into account the
self-interactions of the
particles out of the condensate. This fact  starts improving the ground state energy
results when compared with Bogoliubov theory once $\gamma$
increases ($\gamma > 2$).
 It is very difficult however   for the
Gaussian variational results to get close to the exact for high  $\gamma$
values. In  the one dimensional case when
$\gamma \rightarrow \infty$  the Gaussian ground state energy diverges while the exact goes to the
finite fermi energy.

2) For the particle hole excitations the Gaussian results  take into
account that when we remove a particle from  an occupied state in order to make an excitation, this
particle can be out of the condensate which does not happen in the Bogoliubov scheme. Again
this seems an improved description  for intermediate values of $\gamma$. As $\gamma$ increases
the depletion starts to grow and in this particular case goes to a fermi sea when $\lambda
\rightarrow \infty$ which can not be described by our methods since they do not deal with 
very short range correlations. 

3) Another important point that is somehow related to item 2 is that to solve the RPA equation
we had decoupled harmonic oscillators which , in the quasi-boson picture  means that the RPA
takes into account the interaction between the quasi-bosons generating new modes that are
mixtures of one $(\delta \phi,\delta \pi)$ and two quasi-bosons ($\delta G, \delta \Sigma$).
In the Bogoliubov scheme we always have free quasi-bosons when their
 interaction starts to be relevant as $\gamma$ increases.

We can conclude that the Gaussian variational method can  describe systems where the
depletions can not be neglected (when dilute theories break down) but because of the absence of short
range correlation will require corrections  for highly depleted systems (like Helium 4).

\acknowledgments
We thank the referee for the very useful suggestions. P.T. thanks 
Eddy Timmermans for useful discussions.
This work was supported in part by funds provided by
the U.S. Department of Energy under cooperative agreement
\# DE-FC-94ER 40818. P.T. was supported by Conselho Nacional de
Desenvolvimento Cientifico e Tecnologico (CNPq), Brazil.

\appendix
\section{}
If we carefully examine the expression (\ref{23aa}) we see that the functional integral that must
be computed involve moments of a Gaussian. Namely,
\begin{equation}
\int ({\cal D} \phi') \Psi^{\ast}[\phi',t] \phi'({\bf x}_{1}) \ldots \phi'({\bf x}_{4}) \Psi[
\phi',t]
\end{equation}
\noindent where $\Psi[\phi']$ is our normalized trial wave functional given by (\ref{9}). This 
functional
integrals can be computed easily if we include a source term \cite{BH} in the normalization
integral i.e.
\begin{equation}
\int ({\cal D} \phi') \Psi^{\ast}[\phi',t] e^{\int_{\bf x} J({\bf x}) \delta \phi'({\bf x})}
 \Psi[\phi',t].
\end{equation}
\noindent Using the expression for $\Psi[\phi',t]$ we complete the squares in the exponential
getting
\begin{equation}
\int ({\cal D} \phi') \Psi^{\ast}[\phi',t] e^{\int_{\bf x} J({\bf x}) \delta \phi'({\bf x},t)
}  \Psi[\phi',t] =
e^{\frac{1}{2} \int_{\bf x} J({\bf x}) G({\bf x},{\bf y},t) J({\bf y})}.
\end{equation}
\noindent The source term allows us to compute the functional integral of any moment of the
 Gaussian
\begin{eqnarray}
&& \int ({\cal D} \phi') \delta \phi'({\bf x}_{1}) \ldots \delta \phi'({\bf x}_{n}) e^{-\int_{{\bf x},{\bf
y}}
\delta \phi'({\bf x},t) \frac{G^{-1}({\bf x},{\bf y},t)}{2} \delta \phi'({\bf y},t)} =
\left. \frac{\delta}{\delta J({\bf x}_{1})} \ldots \frac{\delta}{\delta J({\bf x}_{n})}
e^{\frac{1}{2} \int_{\bf x} J({\bf x}) G({\bf x},{\bf y},t) J({\bf y})} \right|_{J=0}.
\end{eqnarray}
\noindent For instance we can calculate
\begin{eqnarray}
\label{apend1}
&& \int ({\cal D} \phi') \Psi^{\ast}[\phi',t]  \phi'({\bf x})  \phi'({\bf y}) V({
\bf x}-{\bf y})
 \phi'({\bf x})  \phi'({\bf y}) \Psi[\phi',t] = 4 \phi({\bf x},t) \phi({\bf y},t) V({\bf x}-{\bf y}) G((
{\bf x},{\bf y},t) \nonumber \\
&&  +
\phi^{2}({\bf x},t)  V({\bf x}-{\bf y}) G(({\bf  y},{\bf y },t) +
\phi^{2}({\bf y },t)  V({\bf x}-{\bf y}) G(({\bf x},{\bf x},t) \nonumber \\
&& 2 G({\bf x},{\bf y},t) V({\bf x}-{\bf y}) G({\bf x},{\bf y},t) + G({\bf x},{\bf x},t) V({
\bf x} - {\bf y},t) G({\bf y},{\bf y},t)
\end{eqnarray}
\noindent Note that the last two terms of (\ref{apend1}) show,as expected, the mean field
factorization.

\section{}
After some algebra one gets the elements of the matrix $W$ 
which appears in equation (\ref{s111}). For calculating the 
dispersion relation, $\Omega$ will be below the lowest 
pole appearing in (B1)-(B9). 

\begin{eqnarray}
W_{1,1} &=& 1 - \lambda \int_{q} \frac{s_{\rm M} x^{2}}{(\Omega^{2} -\Omega_{2}^{2})}-
\lambda^{2} \phi^{2} \int_{q}\frac{ B s_{\rm M} x^{2} + v  x \Omega^{2}
  }{(\Omega^{2} - \Omega_{1}^{2})(\Omega^{2} - \Omega_{2}^{2})} \\
\nonumber \\
W_{1,2} &=& \lambda^{2} \Omega \phi^{2} \int_{q} \frac{A v x + s_{\rm M}
x^{2}}{(\Omega^{2} - \Omega_{1}^{2})(\Omega^{2} - \Omega_{2}^{2})} + \lambda
 \Omega \int_{q} \frac{x y }{(\Omega^{2} - \Omega_{1}^{2})} \\
\nonumber \\
W_{1,3} &=& 2 \lambda^{2} \phi^{2}\Omega \int_{q} \frac{A v x +s_{\rm M}
x^{2} }{(\Omega^{2} - \Omega_{1}^{2})(\Omega^{2} - \Omega_{2}^{2})} +  2
\lambda \Omega \int_{q} \frac{x z}{(\Omega^{2} - \Omega_{2}^{2})} \\
\nonumber \\
W_{2,1} &=&  \lambda \Omega \int_{q} \frac{x  y}{(\Omega^{2} -\Omega_{2}^{2})}
+ \lambda^{2} \phi^{2} \Omega \int_{q} \frac{s_{\rm K} v y + B x y
}{(\Omega^{2} - \Omega_{1}^{2})(\Omega^{2} - \Omega_{2}^{2})} \\
\nonumber \\
W_{2,2} &=& 1 - \lambda^{2} \phi^{2} \int_{q} \frac{A s_{\rm K} v  y
 + x y  \Omega^{2}}{(\Omega^{2} - \Omega_{1}^{2})(\Omega^{2} - \Omega_{2}^{2
})} - \lambda \int_{q} \frac{s_{\rm K} y^{2}}{(\Omega^{2} - \Omega_{1}^{2})} \\
\nonumber \\
W_{2,3} &=& - 2\lambda^{2} \phi^{2} \int_{q} \frac{A s_{\rm K} v y
 + x y  \Omega^{2}}{(\Omega^{2} - \Omega_{1}^{2})(\Omega^{2} - \Omega_{2}^{2}
)} -  2 \lambda \int_{q} \frac{s_{\rm K} y z}{(\Omega^{2} - \Omega_{2}^{2})} \\
 \nonumber \\
W_{3,1} &=& \lambda \Omega \int_{q} \frac{x z}{(\Omega^{2} - \Omega_{2}^{2})} +
  \lambda^{2} \phi^{2} \Omega \int_{ q} \frac{s_{\rm K} v z + B x z
 }{(\Omega^{2} - \Omega_{1}^{2})(\Omega^{2} - \Omega_{2}^{2})} \\
\nonumber \\
W_{3,2} &=& - \lambda^{2} \phi^{2}  \int_{q} \frac{A s_{\rm K} v z
 + x z \Omega^{2}}{(\Omega^{2} - \Omega_{1}^{2})(\Omega^{2} - \Omega_{2}^{2}
)} - \lambda \int_{q} \frac{s_{\rm K} y  z}{(\Omega^{2} - \Omega_{2}^{2})} \\
\nonumber \\
W_{3,3} &=& 1 -  2 \lambda^{2} \phi^{2}  \int_{q} \frac{A s_{\rm K} v z +
 x z \Omega^{2}}{(\Omega^{2} - \Omega_{1}^{2})(\Omega^{2} - \Omega_{2}^{2})}
 -  2 \lambda \int_{q} \frac{s_{\rm K} z^{2}}{(\Omega^{2} - \Omega_{2}^{2})}
\end{eqnarray}

\newpage
{\large Figure Captions}
\vskip 0.25in
FIG.1 The ground state energy per particle can be written as $E/N = \rho^{2}  g(\gamma) $. The 
curves give $ g(\gamma)$ as a function of $\gamma$ for the Gaussian , the Bogoliubov  and  the exact result.
\vskip 0.25in
FIG.2 The exact threshold curves for the particle-hole excitation 
as a function of the total particle - hole momentum $P=k-k'$
as obtained using  (\ref{ee1})
\vskip 0.25in
FIG.3 The  free quasi-boson energy $\Omega_{1}(P)$ given by (\ref{93}). The 
two free quasi-boson 
energy $\Omega_{2}(P,q)$ given by (\ref{94i}) and the new gapless $\Omega(P)$ mode 
 energy obtained with the RPA calculation by solving (\ref{s111}). Note that, as expected, the 
gapless property of $\Omega(P)$ comes from the Goldstone mode associated with the 
symmetry breaking 
\vskip 0.25in
FIG.4 Comparison between the three different particle-hole excitations 
curves: Bogoliubov $\Omega_{1}^{\rm b}(P)$ given by (\ref{bogexc}), the    
RPA $\Omega(P)$ by solving (\ref{s111}) and the  
exact using (\ref{ee1}). These are given as a function of the total particle-hole 
momentum P for $\gamma$ =  0.787 .
\vskip 0.25in
FIG.5 Comparison between the three particle-hole excitations 
curves: $\Omega_{1}^{\rm b}(P)$ given by (\ref{bogexc}), 
$\Omega(P)$ by solving (\ref{s111}) and the 
exact using (\ref{ee1}) as a function of the total particle-hole momentum P for $\gamma$ = 5.527.
\vskip 0.25in

\begin{thebibliography}{99}
\bibitem{FN} In equations
(\ref{64}) for the $G$ equation,  the first delta is a Dirac delta
and the second means an infinitesimal change.
\end{thebibliography}
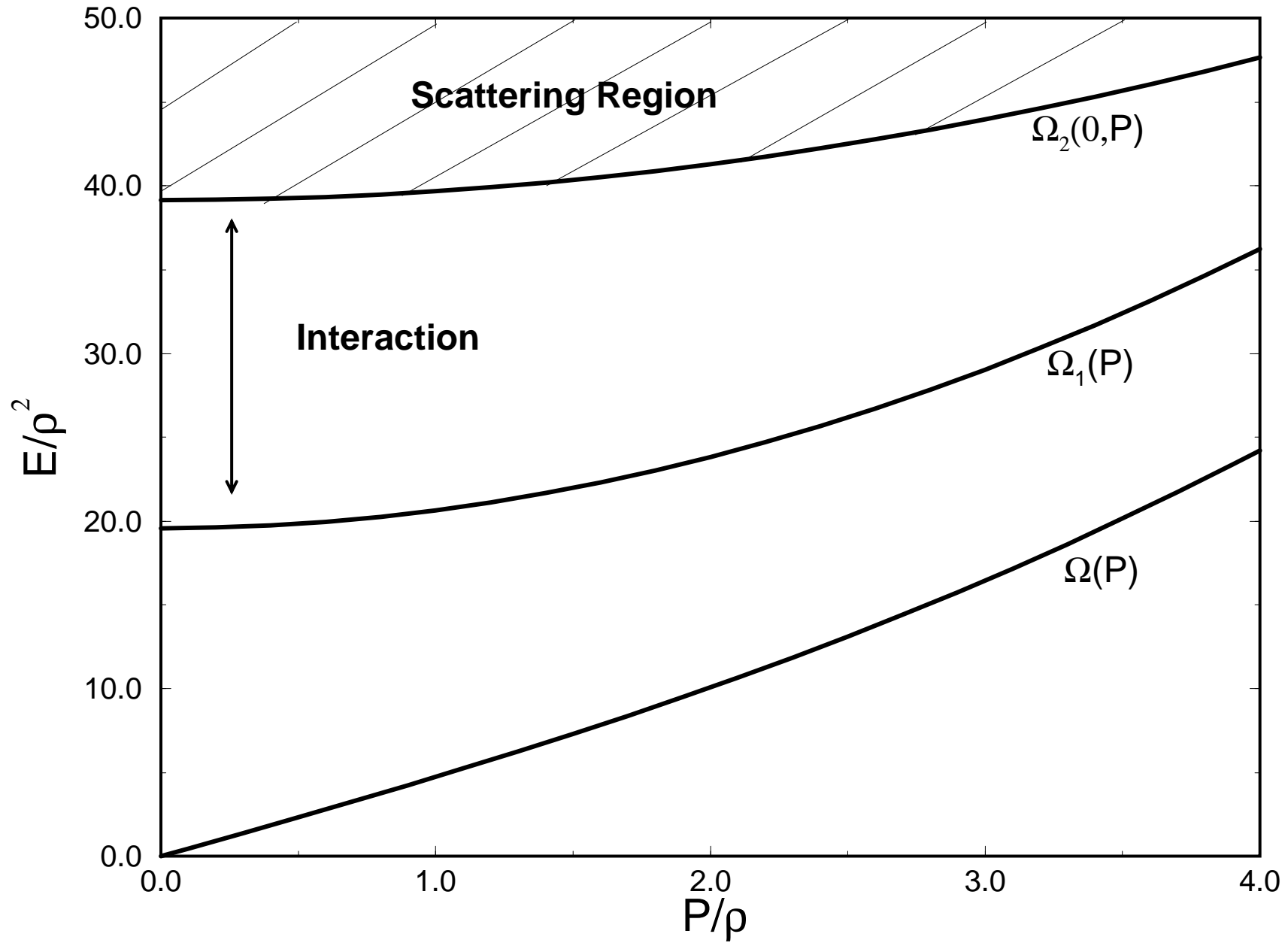
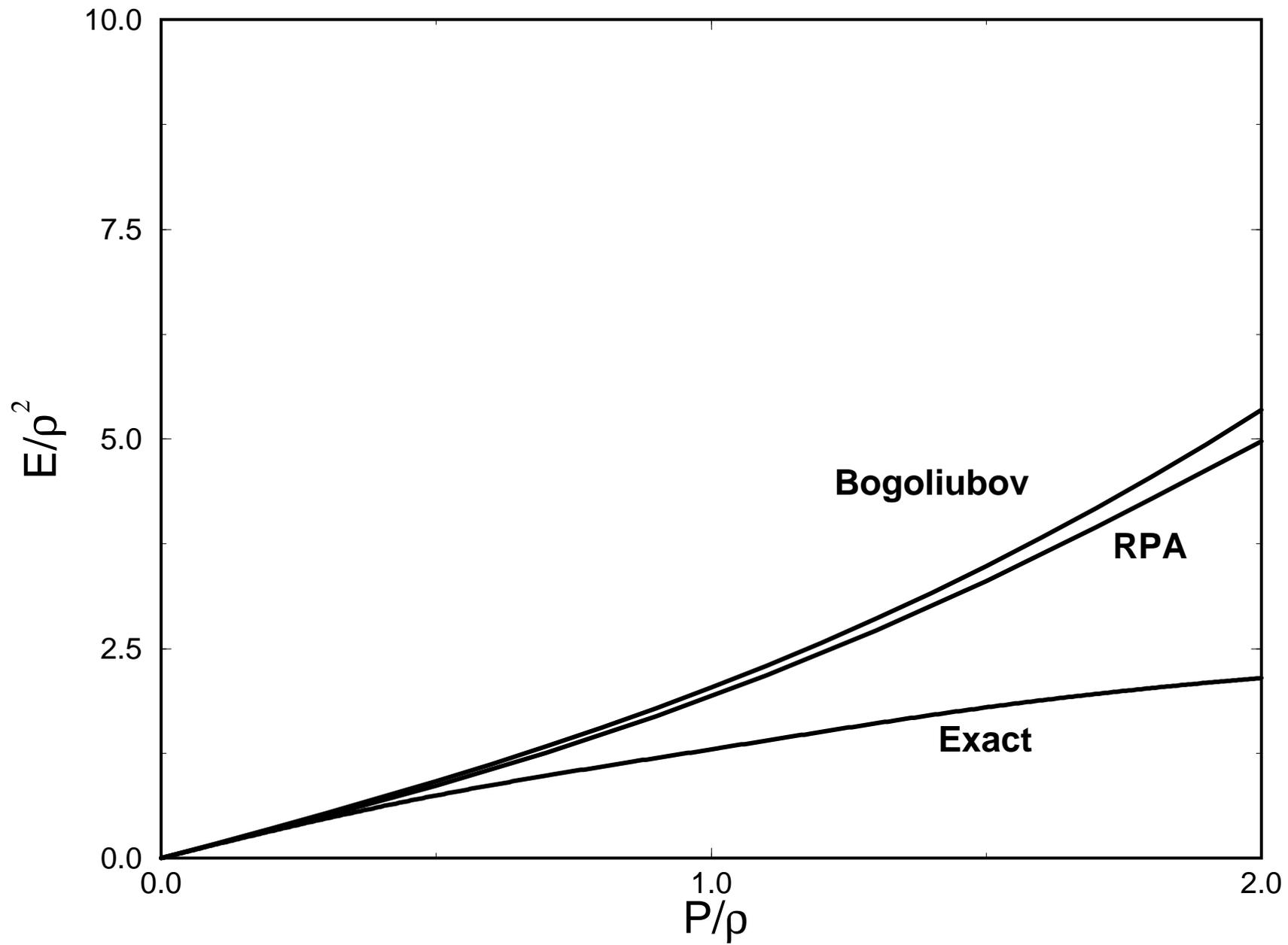
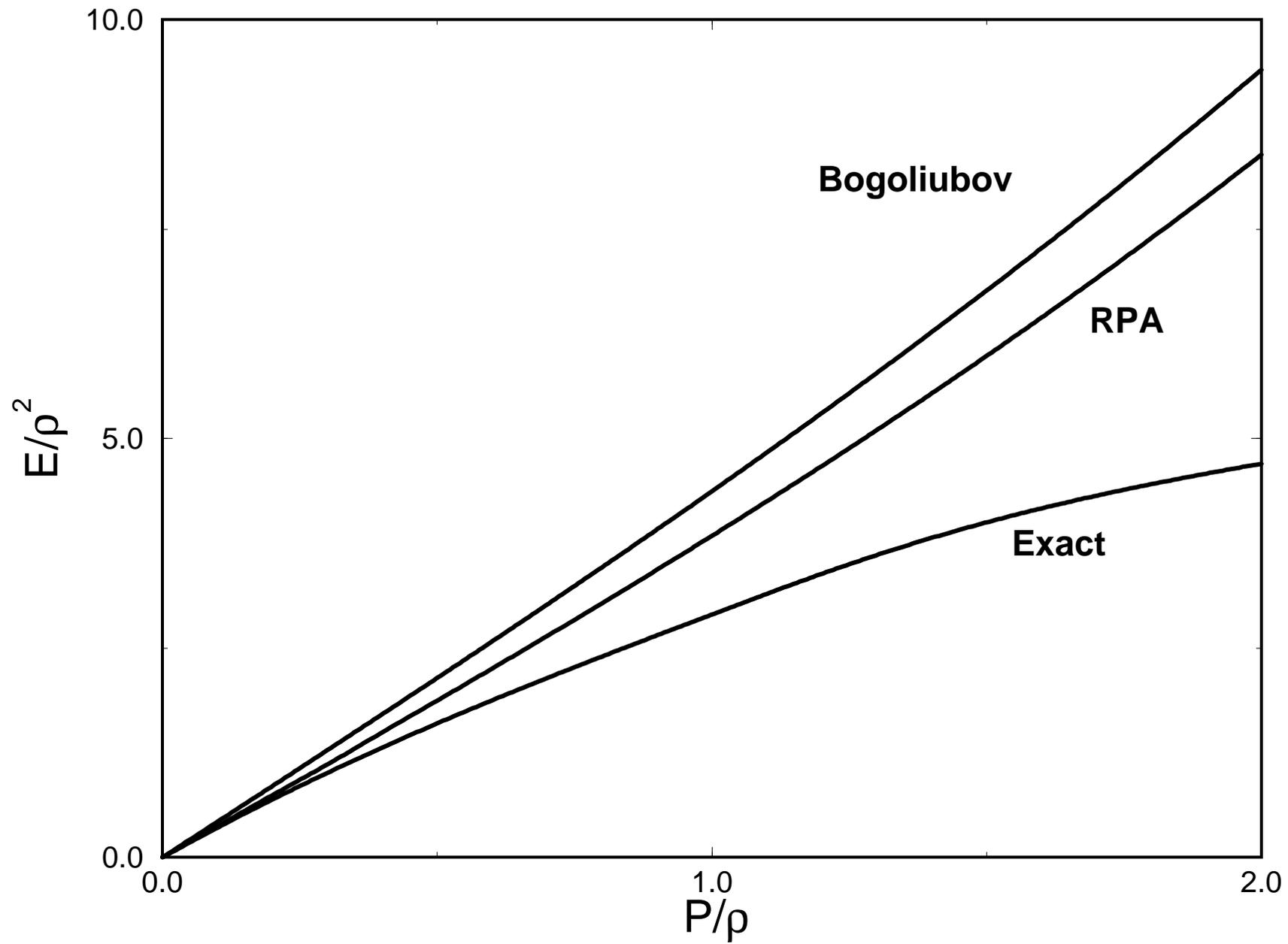
\end{document}